\documentclass[aps,prd,floats,floatfix,amssymb,twocolumn,superscriptaddress,nofootinbib]{revtex4-2}

\usepackage{mathptmx}
\usepackage{physics}
\usepackage{amsmath}
\usepackage{amsfonts,amssymb}
\usepackage{graphicx,graphics}
\usepackage{hyperref}
\usepackage{ulem}
\usepackage{color}
\usepackage{mathrsfs}
\usepackage{bbold}

\allowdisplaybreaks

\begin{document}
\title{Quantum superposition of boundary condition in $\mathrm{PAdS}_2$}

\author{Jo\~ao P. M. Pitelli}
\email[]{pitelli@unicamp.br}
\affiliation{Departamento de Matem\'atica Aplicada, Universidade Estadual de Campinas,
13083-859 Campinas, S\~ao Paulo, Brazil}%

\author{Bruno S. Felipe}
\email[]{brunfeli@ifi.unicamp.br}
\affiliation{Instituto de F\'isica Gleb Wataghin, Universidade Estadual de Campinas, 13083-859 Campinas, S\~ao Paulo, Brazil}

\author{Claudio Dappiaggi}
\email{claudio.dappiaggi@unipv.it}
\affiliation{
Dipartimento di Fisica, Univeristà degli Studi di Pavia \& INFN, Sezione di Pavia, Via Bassi, 6, I-27100 Pavia (PV), Italy}

\author{Elizabeth Winstanley}
\email{e.winstanley@sheffield.ac.uk}
\affiliation{
School of Mathematical and Physical Sciences,
The University of Sheffield,
Hicks Building,
Hounsfield Road,
Sheffield. S3 7RH United Kingdom}

\begin{abstract}
We explore the quantum superposition of boundary conditions in the context of the Poincaré patch of the two-dimensional Anti-de Sitter space ($\mathrm{PAdS}_2$). Focusing on Robin (mixed) boundary conditions (RBC), we investigate the response function of the Unruh-DeWitt (UDW) detector interacting with two or more scalar fields, each respecting a different boundary condition. The role of this quantum superposition is two-fold: i) it may represent different fields propagating on the same spacetime and interacting with an UDW detector or ii) it may describe an UDW detector on a superposition of spacetimes, each one with an inequivalent propagating field.

\end{abstract}

\maketitle

\section{Introduction}

It is known that the vacuum state for quantum fields in Minkowski space is typically constructed to be invariant under the Poincaré group. This means that all inertial observers agree on the state having no particle content. However, this is not true for general curved spacetimes, where the Poincaré group is no longer a symmetry group. In this case, the absence of a ``preferred frame'' leads to a non-unique notion of vacuum. 

Usually, in globally hyperbolic spacetimes having a timelike Killing vector field $\xi$, vacuum states are constructed using the notion of positive frequency (with respect to $\xi$) modes $\phi_j$ satisfying~\cite{birrell}
\begin{equation}
    \mathcal{L}_{\xi}\phi_j=-i\omega_j \phi_j,\,\,\,\omega_j>0,
\end{equation}
where ${\mathcal {L}}_{\xi }$ denotes the Lie derivative.
In non-globally hyperbolic spacetimes, the quantization of fields is more subtle and depends on additional assumptions. In the absence of a Cauchy surface, the evolution of classical fields may not be uniquely determined by initial data on any spacelike surface. However, it was shown in Refs.~\cite{wald,ishibashi-wald} that it is possible to prescribe a sensible evolution for classical scalar fields on a great variety of static non-globally hyperbolic spacetimes through the specification of boundary conditions at the edge of spacetime. This is particularly relevant for static non-globally hyperbolic spacetimes possessing naked singularities~\cite{helliwell,barroso-pitelli} or even a conformal infinity, as in the case of Anti-de Sitter (AdS) spacetime~\cite{pitelli-comment,pitelli-barroso-mosna,ishibashi2}. Consequently, after quantization, the vacuum state becomes dependent on both the timelike Killing field $\xi$ and on the boundary conditions. 

In Ref.~\cite{pitelli-barroso}, one of the authors demonstrated that, for conformal fields adhering to Robin-type boundary conditions at the conformal boundary of $\mathrm{PAdS}_2$, a subtle change between \textit{isometric} frames corresponds to a change in the boundary condition and it results in a finite number of particles being created. 
This raises the following question, which is the target of the present study: how does an observer perceive the interaction with one or more inequivalent (respecting different boundary conditions) scalar fields when traveling through space? Or equivalently, which effects does an observer traveling in a superposition of spacetimes, each one in a different vacuum state (parametrized by a different boundary condition), feel?

To address this question, we will model the observer using an Unruh-DeWitt detector in $\mathrm{PAdS}_2$ with coordinates $(t,z)$, $z>0$, following a static trajectory $z=z_0$. The vacuum will be given by $\ket{0}_{\gamma}$ and the Unruh-DeWitt detector will interact with fields $\phi^{(\gamma_{1})}$, $\phi^{(\gamma_{2})}$, \dots, with the parameters $\gamma_i$ representing the corresponding RBC. We associate to each boundary condition a quantum state $\ket{\gamma_i}$, which acts as a boundary condition selector. This setup enables the detector to interact with a superposition of fields that respect different RBC. 
Furthermore, as we will show below, each parameter $\gamma_i$ corresponds to a specific frame selection, allowing us to interpret the detector's response as the result of interactions with fields in different frames. This method, involving the interaction between the observer (i.e., the detector) and a controlled superposition of states $\ket{\gamma_i}$, has been employed as an operational approach to measure the superposition of spacetimes -- an anticipated effect in quantum theories of gravity (see~\cite{foo1,foo2,foo-ds}).

This paper is organized as follows. In the next section we recover the main results of Ref.~\cite{pitelli-barroso} showing the dependence of the vacuum state on the boundary condition in the context of quantum fields in $\mathrm{PAdS}_2$. Subsequently, in Sec.~\ref{sec:detector}, we discuss how to use the Unruh-DeWitt detector to measure superposition by introducing the boundary condition selector states $\ket{\gamma_i}$. In Sec.~\ref{sec:superposition} we illustrate our main findings by considering the superposition of two different boundary conditions. Our final remarks are presented in Sec.~\ref{sec:conclusion}.

\section{Quantum fields in $\mathrm{PAdS}_2$}

\begin{figure}[!htb]
    \centering
    \includegraphics{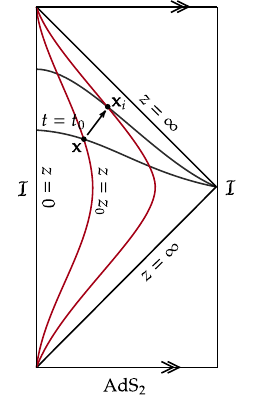}
    \caption{Penrose diagram for $\mathrm{AdS}_2$ spacetime. The Poincaré coordinates ($\mathrm{PAdS}_2$) cover the triangular region from $z=0$ to $z=\infty$. The transformation ${\bf x}\to {\bf x}_i$ is represented as a change between surfaces of constant $t,z$. Note that the $t-$constant spacelike surfaces are not Cauchy surfaces.}
    \label{fig:PAdS}
\end{figure}

The metric of the Poincaré patch of the two-dimensional Anti-de Sitter spacetime is given by 
\begin{equation}
    ds^2=\frac{\ell^2}{z^2}\left(-dt^2+dz^2\right),
\end{equation}
with $t\in \mathbb{R}$ and $z \in \mathbb{R}_{+}$. Here $\ell$ is the AdS curvature radius and, henceforth, we shall set it to $1$. Its conformal structure is presented in Fig.~\ref{fig:PAdS}. As we can observe, in this chart conformal infinity $\mathcal{I}$ corresponds to $z=0$, on which the classical field requires an appropriate boundary condition~\cite{wald,ishibashi-wald,ishibashi2}. For a conformal real scalar field $\varphi:\mathrm{PAdS}_2 \to \mathbb{R}$ satisfying the wave equation $\square \varphi=0$, a suitable general class of boundary conditions at $\mathcal{I}$ is given by the Robin boundary condition given by~\cite{pitelli-tem}
\begin{equation}\label{RBC1}
    \varphi(t,z=0)-\gamma \frac{\partial \varphi(t,z=0)}{\partial z}=0,
\end{equation}
where $\gamma>0$ is a parameter. The particular cases of Dirichlet and Neumann boundary conditions are recovered by setting $\gamma\to 0$ and $\gamma\to\infty$, respectively.

The complete set of positive frequency solutions $\left\{u_{\omega,\gamma}({\bf x})\right\}$ respecting (\ref{RBC1}) and normalized with respect to the Klein-Gordon inner product is given by~\cite{pitelli-barroso}
\begin{equation}
    u_{\omega,\gamma}({\bf x})=\frac{\sin(\omega z)+\gamma \omega \cos(\omega z)}{\sqrt{\pi \omega}\sqrt{1+\gamma^2\omega^2}}e^{-i\omega t}, \quad \omega >0.
\end{equation}
These modes satisfy $\partial_t  u_{\omega,\gamma}=-i\omega  u_{\omega,\gamma}$ and allow us to expand the quantum scalar field as
\begin{equation}
    {\hat {\varphi}}^{(\gamma)}({\bf x})=\int_{0}^{\infty}\mathrm{d}\omega \left( {\hat {a}}_{\omega,\gamma}u_{\omega,\gamma}({\bf x})+{\hat {a}}_{\omega,\gamma}^{\dagger}u_{\omega,\gamma}^{\ast}({\bf x})\right).
\end{equation}
Notice that we explicitly wrote the field dependence on the $\gamma$-parameter. Naturally, the canonical quantization is performed by imposing the usual commutation relation between ${\hat {\varphi}}^{(\gamma)}({\bf x})$ and its conjugated momentum (or equivalently between ${\hat {a}}_{\omega,\gamma}$ and ${\hat {a}}_{\omega,\gamma}^{\dagger}$). Then, the vacuum state $\ket{0}_{\gamma}$ is defined as
\begin{equation}\label{vacuum}
    {\hat {a}}_{\omega,\gamma}\ket{0}_{\gamma}=0,\quad \forall \,\omega >0.
\end{equation}

To understand the effect of the RBC on the vacuum states, consider the coordinate transformation generated by the Killing field  $\xi=t\partial_t+z\partial_z$~\cite{pitelli-barroso} (see the representation in Fig.~\ref{fig:PAdS})
\begin{equation}
    {\bf x}=(t,z)\to {\bf x}_i=(t_i,z_i)=(\lambda_i t, \lambda_i z), \quad \lambda_i >0.
\end{equation}
The RBC (\ref{RBC1}) transforms into
\begin{equation}\label{RBC2}
    \varphi(t_i,z_i=0)-\gamma_i \frac{\partial \varphi(t_i,z_i=0)}{\partial z_i}=0,
\end{equation}
where $\gamma_i=\lambda_i \gamma$ represents a modification of the boundary condition. In other words, if we change the frame, we change the boundary condition. Consequently, we arrive at a new set of mode solutions
\begin{equation}\label{new-modes}
    u_{\widetilde{\omega},\gamma_i}({\bf x})=\frac{\sin(\widetilde{\omega} z)+\gamma_i \widetilde{\omega} \cos(\widetilde{\omega} z)}{\sqrt{\pi \widetilde{\omega}}\sqrt{1+\gamma_i^2\widetilde{\omega}^2}}e^{-i\widetilde{\omega} t},\qquad \widetilde{\omega}>0,
\end{equation}
satisfying (\ref{RBC2}), and the field expanded in this new base, namely ${\hat {\varphi}}^{(\gamma_i)}({\bf x})$, will lead to the definition of a new vacuum state $\ket{0}_{\gamma_i}$ via
\begin{equation}
    {\hat {a}}_{\widetilde{\omega},\gamma_i}\ket{0}_{\gamma_i}=0, \quad \forall \,\widetilde{\omega}>0.
\end{equation}
Therefore, if $\ket{0}_{\gamma}$ denotes the natural vacuum corresponding to the frame in the coordinates ${\bf x}$, then the vacuum for the transformed frame ${\bf x}_i$ is represented by $\ket{0}_{\gamma_i}$. These vacua do not respect $\mathrm{AdS}$ invariance~\cite{pitelli-comment} and depend crucially on the choice of $\gamma_i$. 

The relationship between the two bases of mode solutions was studied in~\cite{pitelli-barroso}, where the authors find the Bogoliubov transformation taking into account the frame change ${\bf x}\to{\bf x}_i$. In this way, the new modes $u_{\widetilde{\omega},\gamma_i}({\bf x})$ can be expressed in terms of the old ones $u_{\widetilde{\omega},\gamma_i}({\bf x})$ as (here, we present the inverse of the transformation in  Eq.~(18) from Ref.~\cite{pitelli-barroso})
\begin{equation}
    u_{\widetilde{\omega},\gamma_i}({\bf x})=\int_{0}^{\infty} \mathrm{d}\omega\left(\alpha_{\omega\widetilde{\omega}}^{\lambda_i}u_{\omega,\gamma}({\bf x})-\beta_{\omega\widetilde{\omega}}^{\lambda_i}u_{\omega,\gamma}^{\ast}({\bf x})\right),
\end{equation}
where $\alpha_{\omega\widetilde{\omega}}^{\lambda_i}$ and $\beta_{\omega\widetilde{\omega}}^{\lambda_i}$ are real Bogoliubov coefficients given by
\begin{widetext}
\begin{equation}\label{bogoliubov}
    \begin{aligned}
        \alpha_{\omega \widetilde{\omega}}^{\lambda_i}&=\frac{1+\lambda_i \gamma^2 \omega^2}{\sqrt{1+\gamma^2\omega^2}\sqrt{1+\gamma^2\lambda_i^2\omega^2}}\delta(\omega-\widetilde{\omega})+\frac{\gamma}{\pi}\frac{\lambda_i-1}{\sqrt{1+\gamma^2\omega^2}\sqrt{1+\gamma^2\lambda_i^2\widetilde{\omega}^2}}\frac{\sqrt{\omega \widetilde{\omega}}}{\omega-\widetilde{\omega}},\\
        \beta_{\omega \widetilde{\omega}}^{\lambda_i}&=\frac{\gamma}{\pi}\frac{\lambda_i-1}{\sqrt{1+\gamma^2\omega^2}\sqrt{1+\gamma^2\lambda_i^2\widetilde{\omega}^2}}\frac{\sqrt{\omega \widetilde{\omega}}}{\omega+\widetilde{\omega}}.
    \end{aligned}
\end{equation}
\end{widetext}
Notably, and for future analysis, we can also express the new annihilation and creation operators as
    \begin{align}
        {\hat {a}}_{\widetilde{\omega},\gamma_i}&=\int_{0}^{\infty} \mathrm{d}\omega \left(\alpha_{\omega\widetilde{\omega}}^{\lambda_i}{\hat {a}}_{\omega,\gamma}+\beta_{\omega\widetilde{\omega}}^{\lambda_i}{\hat {a}}_{\omega,\gamma}^{\dagger}\right) \nonumber \\
        {\hat {a}}_{\widetilde{\omega},\gamma_i}^{\dagger}&=\int_{0}^{\infty} \mathrm{d}\omega \left(\alpha_{\omega\widetilde{\omega}}^{\lambda_i}{\hat {a}}_{\omega,\gamma}^{\dagger}+\beta_{\omega\widetilde{\omega}}^{\lambda_i}{\hat {a}}_{\omega,\gamma}\right).
        \label{ope-transf}
    \end{align}

\section{Unruh-DeWitt detector measuring superposition}\label{sec:detector}

Let us consider the Unruh-DeWitt detector (a two-level system) interacting with a real massless scalar field in $\mathrm{PAdS}_2$. This field respects RBC with parameter $\gamma_i$, where the sub-index $i$ denotes each possible choice for the parameter. Then, the standard interaction Hamiltonian reads\footnote{We do not consider the switching function (responsible for turning the detector on and off) since we are interested in the case of an eternally active detector. Additionally, we are disregarding the internal structure of the detector, which is necessary to obtain a normalized probability.}
\begin{equation}
    \mathcal{H}_{i}=c \sigma(\tau) \varphi^{(\gamma_i)}({\bf x}(\tau)).
\end{equation}
Here, $c$ is a small coupling constant, $\tau$ is the detector's proper time, and $\sigma(\tau)$ is the monopole momentum operator which connects the two-level states of the detector as
\begin{equation}
    \sigma(\tau)=\ket{e}\bra{g}e^{i\Omega \tau}+\ket{g}\bra{e}e^{-i\Omega \tau},
\end{equation}
with $\ket{g}$ and $\ket{e}$ denoting respectively the ground and the excited state of the detector. In the above equation, $\Omega$ is the energy gap between the detector's states. 

Now, in order to describe the superposition of different RBC, we introduce the selector states $\ket{\gamma_i}$ satisfying
\begin{equation}
    \bra{\gamma_i}\ket{\gamma_j}=\delta_{ij}=\left\{\begin{matrix}
1, \quad i=j\\ 
0, \quad i\neq j
\end{matrix}\right.,
\end{equation}
which can be understood as a {\it boundary condition selector} -- or equivalently, a {\it frame selector} -- which will couple the detector with a specified field respecting the boundary condition assigned by $\gamma_i$. In this way, we can write the total interaction Hamiltonian between the detector and $N$ scalar fields (respecting $N$ different RBC) as
\begin{equation}
    \mathcal{H}_{\text{int}}=\sum_{i=1}^{N} \mathcal{H}_{i}\otimes \ket{\gamma_i}\bra{\gamma_i}.
\end{equation}

First of all, we set the vacuum in the observer's (detector's) frame as $\ket{0}_{\gamma}$, i.e., the vacuum for $\varphi^{(\gamma)}({\bf x})$ which satisfies the boundary condition (\ref{RBC1}). Then, the initial state of the total system can be represented as
\begin{equation}
    \ket{\text{in}}= \ket{g}\otimes \ket{0}_{\gamma}\otimes \ket{\text{S}_{\text{in}}}, \,\,\text{with}\,\, \ket{\text{S}_{\text{in}}}=\frac{1}{\sqrt{N}}\sum_{i=1}^{N}\ket{\gamma_i},
\end{equation}
where $\ket{S_{\text{in}}}$ denotes the superposition of $N$ possible boundary conditions. 

After the interaction, we expect the detector and the field to exchange energy $\Omega$ and evolve to their respective excited states. Consequently, the final state of the total system can be written as

\begin{equation}
    \ket{\text{out}}=\ket{e}\otimes \ket{\psi} \otimes \ket{\text{S}_{\text{out}}},
\end{equation}
where $\ket{\psi}={\hat {a}}_{\omega,\gamma}^{\dagger}\ket{0}_{\gamma}$ is the first excited state of the field and $\ket{\text{S}_{\text{out}}}$ characterizes the final superposed boundary conditions. We can assume a controlled superposition by selecting a specific combination of $\ket{\gamma_i}$~\cite{foo1}. Alternatively, a reasonable assumption is that the interaction may cause the individual control states $\ket{\gamma_i}$ to evolve separately by some relative phase.
In other words, each control state acquires a relative phase associated with its respective boundary condition, so that
\begin{equation}
    \ket{\text{S}_{\text{out}}}=\frac{1}{\sqrt{N}}\sum_{i=1}^{N}e^{-i\theta_i}\ket{\gamma_i}.
\end{equation}

By applying perturbation theory, the probability amplitude for the interaction that induces transitions from $\ket{\text{in}}$ to $\ket{\text{out}}$ is 
\begin{equation}
    \begin{aligned}
        &\mathcal{A}_{\text{in}\to\text{out}}=\bra{\text{out}}\ket{\text{in}}-i\bra{\text{out}}\int_{-\infty}^{\infty}\mathrm{d} \tau \, \mathcal{H}_{\text{int}}\ket{\text{in}}+\mathcal{O}(c^2)\\
        &=-\frac{ic}{N} \sum_{i=1}^{N}e^{i \theta_i}\int_{-\infty}^{\infty}\mathrm{d}\tau \, e^{i \Omega \tau} \bra{\psi}\varphi^{(\gamma_i)}({\bf x}(\tau))\ket{0}_{\gamma}+\mathcal{O}(c^2).
    \end{aligned}
\end{equation}
Thus, by squaring the modulus of $\mathcal{A}$ and summing over all final states $\psi$, we derive the non-normalized transition probability (to first order in $c$) as
\begin{equation}\label{probability}
    \mathcal{P}(\Omega)=\frac{c^2}{N^2}\left(\sum_{i=1}^{N}\mathcal{F}^{ii}(\Omega)+\sum_{i\neq j}^{N}e^{-i\phi_{ji}}\mathcal{F}^{ji}(\Omega)\right),
\end{equation}
where $\phi_{ji}:=\theta_j-\theta_i$ and  the {\it $ji$-response function} $\mathcal{F}^{ji}$ is given by
\begin{equation}\label{response}
    \mathcal{F}^{ji}(\Omega)=\int_{-\infty}^{\infty} \mathrm{d}\tau  \, e^{-i \Omega \tau}\int_{-\infty}^{\infty} \mathrm{d}\tau^{\prime} \, e^{i \Omega \tau^{\prime}}\mathcal{W}^{ji}({\bf x},{\bf x}^{\prime}),
\end{equation}
with $\mathcal{W}^{ji}({\bf x},{\bf x}^{\prime})={}_{\gamma}\bra{0}{\hat {\varphi }}^{(\gamma_j)}({\bf x}(\tau)){\hat {\varphi }}^{(\gamma_i)}({\bf x}(\tau^{\prime}))\ket{0}_{\gamma}$ being the two-point function for the fields defined by boundary parameters $\gamma_i$ and $\gamma_j$, acting on the vacuum defined by the mass parameter $\gamma$.

Identifying $\gamma_i$ as the transformed boundary condition (\ref{RBC2}), which implies that $\lambda_i=\gamma_i/\gamma$, we expand the fields ${\hat {\varphi }}^{(\gamma_i)}({\bf x}(\tau))$ in terms of their corresponding modes $u_{\widetilde{\omega},\gamma_i}({\bf x})$ and use the transformation (\ref{ope-transf}) to find
\begin{widetext}
\begin{equation}\label{two-point}
    \begin{aligned}
        \mathcal{W}^{ji}({\bf x},{\bf x}^{\prime})&=\int \mathrm{d}\widetilde{\omega}\int \mathrm{d}\widetilde{k}\int \mathrm{d}\omega\left(\alpha_{\omega \widetilde{\omega}}^{\lambda_j}\beta_{\omega \widetilde{k}}^{\lambda_i}u_{\widetilde{\omega},\gamma_j}({\bf x})u_{\widetilde{k},\gamma_i}({\bf x}^{\prime})+\alpha_{\omega \widetilde{\omega}}^{\lambda_j}\alpha_{\omega\widetilde{k}}^{\lambda_i}u_{\widetilde{\omega},\gamma_j}({\bf x}^{\prime})u_{\widetilde{k},\gamma_i}^{\ast}({\bf x}^{\prime})\right.\\ &\qquad\qquad\qquad\qquad+\left.\beta_{\omega\widetilde{\omega}}^{\lambda_j}\beta_{\omega\widetilde{k}}^{\lambda_i}u_{\widetilde{\omega},\gamma_j}^{\ast}({\bf x})u_{\widetilde{k},\gamma_i}({\bf x}^{\prime})+\beta_{\omega \widetilde{\omega}}^{\lambda_j}\alpha_{\omega\widetilde{k}}^{\lambda_i}u_{\widetilde{\omega},\gamma_j}^{\ast}({\bf x})u_{\widetilde{k},\gamma_i}^{\ast}({\bf x}^{\prime})\right).
    \end{aligned}
\end{equation}
\end{widetext}
As we can observe in (\ref{bogoliubov}), if $\lambda_i = \lambda_j = 1$, Eq.~(\ref{two-point}) simplifies to
\begin{equation}
    \mathcal{W}({\bf x},{\bf x}^{\prime})=\int_{0}^{\infty} \mathrm{d}\omega\, u_{\omega,\gamma}({\bf x}) u_{\omega,\gamma}^{\ast}({\bf x}^{\prime}),
\end{equation}
which is simply the Wightman function expressed as a sum of the modes.

\subsection{Critical acceleration}
We are interested in the simple trajectory where the detector remains at rest at a constant position $z_0$ and evolves only in time (see Fig.~\ref{fig:PAdS}). In terms of the detector's proper time, this trajectory is described by
\begin{equation}\label{trajectory}
    {\bf x}(\tau)=\left(z_0\tau,z_0\right).
\end{equation}
This path is an accelerated trajectory with proper acceleration
\begin{equation}
    a=\sqrt{a^{\mu}a_{\mu}}=\ell=1,
\end{equation}
recalling that $\ell$ is the radius of curvature of $\mathrm{AdS}$. As shown in~\cite{jennings}, $a=\ell$ characterizes a critical acceleration, as thermal effects occur only for $a>\ell$. Since no response is expected from a standard detector following the trajectory (\ref{trajectory}), any non-zero response observed by our approach will indicate the presence of a superposition effect.
 
In this trajectory, the integration over $\tau$ and $\tau^{\prime}$ in the response function (\ref{response}) simplifies to the Fourier transform of the modes $u_{\widetilde{\omega},\gamma_i}$ and their conjugates. Specifically, we have
\begin{subequations}
\begin{align}
        \int_{-\infty}^{\infty} \mathrm{d}\tau \, e^{-i\Omega \tau} u_{\widetilde{\omega},\gamma_j}(z_0 \tau, z_0)&=u_{\widetilde{\omega},\gamma_j}(0,z_0)\delta (\Omega+\widetilde{\omega}z_0),\label{termo1}\\
        \int_{-\infty}^{\infty} \mathrm{d}\tau  \, e^{-i\Omega \tau} u_{\widetilde{\omega},\gamma_j}^{\ast}(z_0 \tau, z_0)&=u_{\widetilde{\omega},\gamma_j}^{\ast}(0,z_0)\delta (\Omega-\widetilde{\omega}z_0),\label{termo2}\\
        \int_{-\infty}^{\infty} \mathrm{d}\tau^{\prime} \, e^{i\Omega \tau^{\prime}} u_{\widetilde{k},\gamma_i}(z_0 \tau^{\prime}, z_0)&=u_{\widetilde{k},\gamma_i}(0,z_0)\delta (\Omega-\widetilde{k}z_0),\label{termo3}\\
        \int_{-\infty}^{\infty} \mathrm{d}\tau^{\prime} \, e^{i\Omega \tau} u_{\widetilde{k},\gamma_i}^{\ast}(z_0 \tau^{\prime}, z_0)&=u_{\widetilde{k},\gamma_i}^{\ast}(0,z_0)\delta (\Omega+\widetilde{k}z_0).\label{termo4}
\end{align}
\end{subequations}
As we can observe, the Dirac delta function enforces the interaction to select only the modes with energy $\pm \widetilde{\omega}_{0}=\pm\Omega/z_0$. 
If $\Omega>0$ (excitation), only the combination of the terms (\ref{termo2}) and (\ref{termo3}) will produce non-zero contributions to the two-point function (\ref{two-point}). Conversely, for $\Omega<0$ (de-excitation), the non-zero contribution arises from the combination of terms (\ref{termo1}) and (\ref{termo4}). As a result, we obtain the $ji$-response functions
\begin{equation}\label{resp-positivo}
    \mathcal{F}^{ji}(\Omega)=u_{{\widetilde {\omega }}_{0},\gamma_j}^{\ast}(0,z_0)u_{{\widetilde {\omega }}_{0},\gamma_i}(0,z_0)\int \mathrm{d}\omega  \,\beta_{\omega {\widetilde {\omega }}_{0}}^{\lambda_j}\beta_{\omega {\widetilde {\omega }}_{0}}^{\lambda_i},
\end{equation}
for $\Omega>0$, and
\begin{multline}\label{resp-negativo}
    \mathcal{F}^{ji}(\Omega)=u_{-{\widetilde {\omega }}_{0},\gamma_j}(0,z_0)u_{-{\widetilde {\omega }}_{0},\gamma_i}^{\ast}(0,z_0)
    \\ \times
    \int \mathrm{d}\omega \,\alpha_{\omega (-{\widetilde {\omega }}_{0})}^{\lambda_j}\alpha_{\omega (-{\widetilde {\omega }}_{0})}^{\lambda_i},
\end{multline}
for $\Omega<0$.

Note that the coefficients $\alpha^{\lambda_i}_{\omega,\widetilde{\omega}}$ (\ref{bogoliubov}) depend on the Dirac delta function, which makes the integrand in (\ref{resp-negativo}) dependent on $\delta^2\left(\omega+\Omega/z_0\right)$, and therefore, the integral is ill-posed. In fact, terms of the form $|\alpha_{\omega \widetilde{\omega}}^{\lambda_i}|^2$ are related to the number of particles only for non-vacuum states. This is due to the fact that quantization of fields with $\gamma_i\neq \gamma_j$ leads to unitarily inequivalent representations. However, this problem is suppressed for an infinite time interaction between the detector and the fields. Hence, we only consider the excitation probability for eternally active detectors. 

\subsection{Detector excitation ($\Omega>0$)}

For $\Omega>0$ the $ji$-response function can be found analytically. Note that the integration in Eq.~(\ref{resp-positivo}) depends only on the coefficients $\beta^{\lambda_i}_{\omega\widetilde{\omega}_{0}}$ and $\beta^{\lambda_j}_{\omega\widetilde{\omega}_{0}}$. This means the integration measures the projection of the modes $u_{\widetilde{\omega}_{0},\gamma_i}$ and $u_{\widetilde{\omega}_{0},\gamma_j}$ into the vacuum $\ket{0}_{\gamma}$. By expressing the modes in (\ref{resp-positivo}) using (\ref{new-modes}) and performing the integration with the explicit expression (\ref{bogoliubov}) for the Bogoliubov coefficients, we arrive at
    \begin{equation}
    \mathcal{F}^{ji}(\Omega)=\frac{\left(\sin\Omega+\gamma_j \Omega \cos \Omega\right)\left(\sin\Omega+\gamma_i \Omega \cos \Omega\right)}{\pi \frac{\Omega}{z_0} \sqrt{\left(\Omega ^2 \gamma _i^2+z_0^2\right) \left(\Omega ^2 \gamma _j^2+z_0^2\right)}} N^{ji}_{\gamma}(\Omega),
\end{equation}
where
\begin{widetext}
\begin{equation}\label{number}
    N^{ji}_{\gamma}(\Omega)=\int_{0}^{\infty} \mathrm{d}\omega \, \beta_{\omega {\widetilde {\omega }}_{0}}^{\lambda_j}\beta_{\omega {\widetilde {\omega }}_{0}}^{\lambda_i}=\frac{(\gamma_j-\gamma)(\gamma_i-\gamma)\left[\pi \gamma \Omega z_0-\gamma^2\Omega^2-\left(z_0^2-\gamma^2\Omega^2\right)\ln \left(\frac{\gamma \Omega}{z_0}\right)-z_0^2\right]}{\pi^2(z_0^2+\gamma^2\Omega^2)^2\sqrt{(z_0^2+\gamma_j^2\Omega^2)(z_0^2+\gamma_i^2\Omega^2)}}.
\end{equation}
\end{widetext}

First, note that we find a finite $ji$-response function for the detector that is always switched on, and consequently, a finite probability transition. This is a different result from particle production along an accelerated path, which is usually infinite (even in Minkowski space). Furthermore, for $\gamma_j=\gamma_i=\gamma$ (or equivalently $\lambda_j=\lambda_i=1$), we obtain $N^{ji}_\gamma(\Omega)=0$, as expected for non-superposition on such a trajectory.

\begin{figure}[!htb]
    \centering
    \includegraphics[scale=0.45]{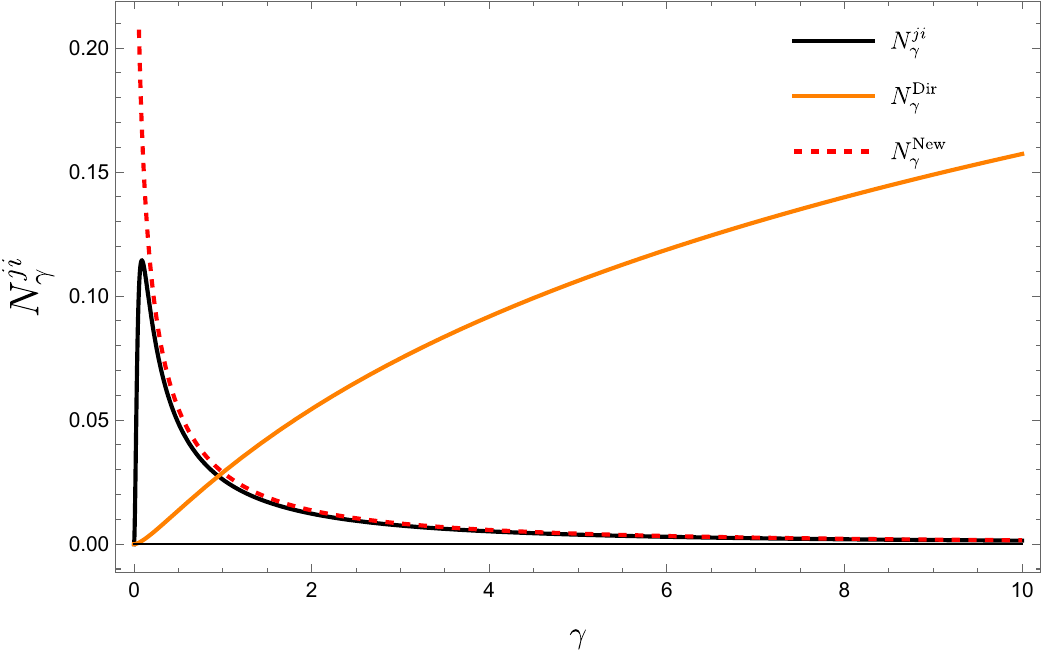}
    \caption{Illustration of $N^{ji}_{\gamma}$ (\ref{number}) as a function of the boundary condition parameter $\gamma$. For $N^{ji}_{\gamma}$ we take $\lambda_i=\lambda_j=20$ and for all curves we are setting $\Omega=z_0=1$. Here, $N^{\text{Dir}}_{\gamma}$ ($N^{\text{Neu}}_{\gamma}$) is the projection of the modes $u_{\widetilde{\omega}_{0},0}$ ($u_{\widetilde{\omega}_{0},\infty}$) onto the vacuum $\ket{0}_{\gamma}$.}
    \label{fig:N}
\end{figure}

By analyzing Eq.~(\ref{number}), we can explore the extreme cases of the initial field in the detector's frame by setting specific values for $\gamma$. For the limits $\gamma \to 0$ and $\gamma \to \infty$, while keeping $\lambda_i$ and $\lambda_j$ constant, we find $N^{ji}_{0}=N^{ji}_{\infty}=0$, i.e., the Dirichlet vacuum $\ket{0}_{0}$ and Neumann vacuum $\ket{0}_{\infty}$ do not contain Robin modes $u_{\widetilde{\omega}_{0},\gamma_i}$ and $u_{\widetilde{\omega}_{0},\gamma_j}$. On the other hand, we can consider the Robin vacuum to interact with Dirichlet and Neumann modes by taking limits for $\gamma_i$ and $\gamma_j$. Writing
\begin{equation}
N^{\text{Dir}}_{\gamma} = \lim_{\gamma_i \to 0} N^{ii}_{\gamma},\quad N^{\text{Neu}}_{\gamma} = \lim_{\gamma_i \to \infty} N^{ii}_{\gamma},
\end{equation}
we illustrate these limits in Fig.~\ref{fig:N}. As can be observed, the transformation $\gamma_i=\lambda_i\gamma$ combines $\gamma$ and $\gamma_i$ in a way that always yields a finite projection of modes dependent on $\gamma_i$ in the vacuum $\ket{0}_{\gamma}$. Nonetheless, if we fix the modes to respect the Dirichlet (Neumann) boundary condition, the projection onto the Neumann (Dirichlet) vacuum goes to infinity. 
We also see from Fig.~\ref{fig:N} that $N^{ji}_{\gamma}$ is very similar to the Neumann limit for all values of $\gamma $ except those close to zero.

\section{Example of superposition for $N=2$}\label{sec:superposition}

The simplest non-zero $ii$-response function occurs when we consider the control states which select only one boundary condition, i.e., $\ket{\text{S}_{\text{out}}}\equiv\ket{\gamma_1}$ (equivalently $N=1$). In this case, $\mathcal{F}^{11}(\Omega)$ represents how the detector perceives the particles from the modes $u_{\widetilde{\omega}_{0},\gamma_1}$ in the vacuum state $\ket{0}_{\gamma}$, analogous to the study in Ref.~\cite{pitelli-barroso} on the observer's perception of the ``subtle frame change.''

\begin{figure}[!htb]
    \centering
    \includegraphics[scale=0.45]{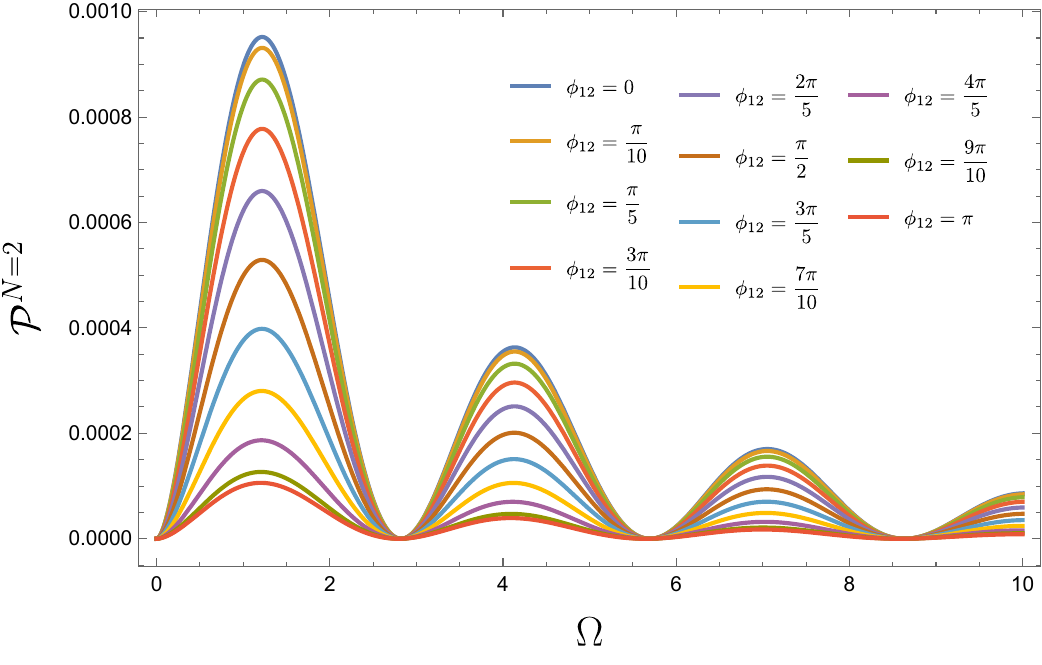}
    \caption{Probability $\mathcal{P}(\Omega)$ as a function of $\Omega$ for $N=2$, setting the coupling $c=1$. We also fix $z_0=10$, $\gamma=1$, $\lambda_1=1.1$, $\lambda_2=1.2$ and vary $\phi_{12}$ from $0$ to $\pi$.}
    \label{fig:N=2-energy}
\end{figure}

For $N=2$, we encounter the first superposition of two different Robin boundary conditions. The probability (\ref{probability}) simplifies to
\begin{equation}
    \mathcal{P}(\Omega)=\frac{c^2}{4}\left[\mathcal{F}^{11}(\Omega)+\mathcal{F}^{22}(\Omega)+2\mathcal{F}^{12}(\Omega)\cos \phi_{12}\right],
\end{equation}
illustrated for particular values in Fig.~\ref{fig:N=2-energy}. This figure demonstrates the probability's oscillatory dependence on $\Omega$. By considering the equation $\mathrm{d}\mathcal{P}(\Omega)/\mathrm{d}\Omega=0$, we can find energies $\Omega_0$ where the minimum values of $\mathcal{P}(\Omega)$ are reached, indicating that the detector does not interpret the superposition with these energies as particles. These values can be determined numerically. For the values considered in Fig.~\ref{fig:N=2-energy} we find
\begin{equation}
    \Omega_0\approx \left\{\begin{matrix}
 2.82,\,5.69,\,8.63,\,\dots& \text{for}\,\,\phi_{12}=0, \\ 
 2.79,\,5.64,\, 8.57,\,\dots & \text{for}\,\,\phi_{12}=\pi.
\end{matrix}\right.
\end{equation}
These values are very similar, suggesting that superposition has minimal influence on the energies which are unobserved by the detector. 
Also in Fig.~\ref{fig:N=2-energy}, we see that successive oscillations are damped as $\Omega $ increases.
The amplitude of the oscillations also decreases as $\phi _{12}$ increases. 

\begin{figure}[!tb]
    \centering
    \includegraphics[scale=0.45]{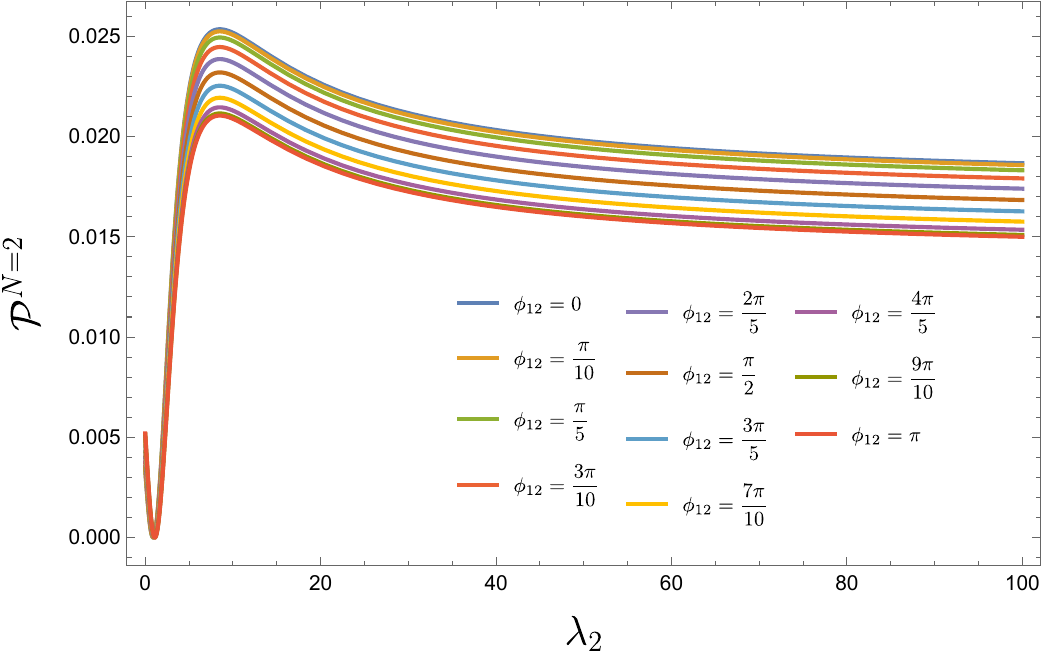}
    \caption{Probability $\mathcal{P}(\Omega)$ as a function of $\lambda_2$ for $N=2$ and coupling $c=1$. We set $\Omega=1$, $z_0=3$ and $\lambda_1=1.1$ and show  results for $\phi_{12}$ varying from $0$ to $\pi$.}
    \label{fig:N=2-lambda}
\end{figure}

We also investigate the dependence of $\mathcal{P}(\Omega)$ on the parameters $\lambda_i$. For $N=2$ and $\lambda_1=1.1$, we obtain Fig.~\ref{fig:N=2-lambda}. As observed, the probability decreases for $0<\lambda_2<\lambda_{\text{min}}$ and increases for $\lambda_{\text{min}}<\lambda_2<8.5$, where $\lambda_{\text{min}}$ depends on the phase difference $\phi_{12}$. For $\phi_{12}=0,\pi$ we find $\lambda_{\text{min}}=0.9,\,1.1$, respectively. Furthermore, as the rate $\lambda_2$ approaches infinity, the probability approaches a constant value dependent on $\phi_{12}$
\begin{equation}
    \mathcal{P}^{N=2}(\Omega)\xrightarrow{\lambda_2\to\infty}0.0157+0.0017\cos{\phi_{12}},
\end{equation}
for the values considered in Fig.~\ref{fig:n=2-lambda}.

\section{Conclusion}\label{sec:conclusion}

In this work, we investigated the quantum superposition of Robin boundary conditions for scalar fields in $\mathrm{PAdS}_2$ spacetime. In particular, we  analyzed the response function of an Unruh-DeWitt detector interacting with two or more scalar fields with different boundary conditions. This approach has two significant interpretations: to explore scenarios where these multiple inequivalent fields coexist, or where the spacetime itself exhibits quantum superposition effects.

We found that when the detector is in the Robin vacuum state with $\gamma>0$ and interacts with any other field with $\tilde{\gamma}\neq \gamma$, the response function is nonzero, but remains finite. In contrast, a detector in Dirichlet or Neumann vacuum states is completely blind to any other field with $\gamma>0$. 

As a specific case, we examined the superposition of two different boundary conditions, $\gamma_1$ and $\gamma_2$. We observed that, depending on the relative phase $\phi_{12}$ (i.e., on the final superposition state $\ket{\text{S}_{\text{out}}}$), certain energy values $\Omega_0$ do not excite the detector. 
Furthermore, as the rate $\lambda_2$ increases, the transition probability converges to a constant value that is determined by $\phi_{12}$.

\acknowledgments

J. P. M. P. thanks the support provided in part by Conselho Nacional de Desenvolvimento Científico e Tecnológico (CNPq, Brazil), Grant No. 311443/2021-4, and Fundação de Amparo à Pesquisa do Estado de São Paulo (FAPESP) Grant No. 2022/07958-4. B. S. F. acknowledges support from the Conselho Nacional de Desenvolvimento Científico e Tecnológico (CNPq, Brazil), Grant No. 161493/2021-1.  The work of E.W.~is supported by STFC grant number ST/X000621/1.

\end{document}